\documentclass[preprint,12pt]{elsarticle}

\usepackage{amssymb}
\usepackage{amsmath}

\usepackage{dsfont}

\newcommand{\dd}{\mathrm{d}}
\newcommand{\ee}{\mathrm{e}}
\newcommand{\ii}{\mathrm{i}}
\newcommand{\R}{\mathds R}
\newcommand{\CC}{\mathds C}

\newcommand{\E}{\mathcal E}

\newcommand{\A}{\mathcal A}
\renewcommand{\H}{{\mathcal H}}
\newcommand{\C}{{\mathcal C}}

\newcommand{\Y}{{\mathbf Y}}
\newcommand{\J}{{\mathbf J}}
\newcommand{\B}{{\mathbf B}}

\newcommand{\1}{{\mathbf 1}}
\newcommand{\0}{{\mathbf 0}}
\newcommand{\EE}{{\mathbf E}}
\renewcommand{\Re}{\mathrm{Re}}
\renewcommand{\Im}{\mathrm{Im}}

\newcommand{\Cmat}{{\mathbf C}}
\newcommand{\Dmat}{{\mathbf D}}

\journal{Wave Motion}
\begin{document}

\begin{frontmatter}


\title{Soliton methods and the black hole balance problem}
\author{J\"org Hennig}
\ead{joerg.hennig@otago.ac.nz}
\address{Department of Mathematics and Statistics, University of Otago,\\
PO Box 56, Dunedin 9054, New Zealand}

\begin{abstract}

This article is an extended version of a presentation given at KOZWaves 2024: The 6th Australasian Conference on Wave Science, held in Dunedin, New Zealand.

Soliton methods were initially introduced to study equations such as the Korteweg--de Vries equation, which describes nonlinear water waves. Interestingly, the same methods can also be used to analyse equilibrium configurations in general relativity. An intriguing open problem is whether a relativistic $n$-body system can be in stationary equilibrium. 
Due to the nonlinear effect of spin-spin repulsion of rotating objects, and possibly considering charged bodies with additional electromagnetic repulsion, the existence of such unusual configurations remains a possibility.
An important example is a (hypothetical) equilibrium configuration with $n$ aligned black holes. 
By studying a linear matrix problem equivalent to the Einstein equations for axisymmetric and stationary (electro-) vacuum spacetimes, we derive the most general form of the boundary data on the symmetry axis in terms of a finite number of parameters.
In the simplest case $n=1$, this leads to a constructive uniqueness proof of the Kerr (--Newman) solution. For $n=2$ and vacuum, we obtain non-existence of stationary two-black-hole configurations. For $n=2$ with electrovacuum, and for larger $n$, it remains an open problem whether the well-defined finite solution families contain any physically reasonable solutions, i.e.\  spacetimes without anomalies such as naked singularities, magnetic monopoles, and struts.

\end{abstract}



\end{frontmatter}

\section{Introduction}
\label{sec:Intro}
\subsection{Solitons}

Wave phenomena governed by \emph{linear} wave equations are usually subject to dispersion and dissipation. Solitons, on the other hand, are remarkable \emph{nonlinear} waves that preserve their shape, are stable, and recover even after collisions with other solitons. Probably the first documented observation of an approximate soliton in nature was made in 1834 by Scottish engineer and shipbuilder John Scott Russel. Later, he described what he called the `wave of translation' as follows \cite{Russel1845}.
\begin{quote}\it
 I was observing the motion of a boat which was rapidly drawn along a narrow channel by a pair of horses, when the boat suddenly stopped --- not so the mass of water in the channel which it had put in motion; 
it accumulated round the prow of the vessel in a state of violent agitation, then suddenly leaving it behind, rolled forward with great velocity, assuming the form of a large solitary elevation, a rounded, smooth and well-defined heap of water, which continued its course along the channel apparently without change of form or diminution of speed. I followed it on horseback, and overtook it still rolling on at a rate of some eight or nine miles an hour, preserving its original figure some thirty feet long and a foot to a foot and a half in height. Its height gradually diminished, and after a chase of one or two miles I lost it in the windings of the channel. Such, in the month of August 1834, was my first chance interview with that singular and beautiful phenomenon which I have called the Wave of Translation~[\dots].
\end{quote}

An appropriate mathematical description of waves on shallow%
\footnote{The derivation of the KdV equation includes a Taylor expansion of the fluid equations about the channel bed. Hence, the resulting equation is  an approximation that is only valid in regions of shallow water depth.} 
water surfaces is provided by the famous Korteweg--de Vries (KdV) equation, which was first derived in 1877 by Boussinesq \cite{Boussinesq1877} and later rediscovered in 1895 by its namesakes, Korteweg and de Vries \cite{KdV1895}. In terms of dimensionless variables, the KdV equation for the amplitude $u(x,t)$  of a wave propagating in $x$-direction reads
\begin{equation}\label{eq:KdV}
 u_{,t}+u_{{,xxx}}+6u u_{,x}=0,
\end{equation}
where a comma denotes partial derivatives with respect to the indicated variables.
Korteweg and de Vries also constructed the corresponding single-soliton solution $u=2\kappa^2/\cosh^2(\kappa x-4\kappa^3 t-c)$, where  $\kappa$ and $c$ are constants. This describes a travelling wave exactly as observed by Russel. Further exact solutions with multiple solitons have later been obtained, and finally the complete initial-value problem for the KdV equation was solved in 1967 by Gardner, Greene, Kruskal, and Miura~\cite{Gardner1967}.

Besides the KdV equation, there are many other `soliton equations', for which exact solutions can be constructed. Some examples are:
\begin{itemize}
 \item  The {\bf Sine--Gordon equation} 
    \begin{equation}
     \varphi_{,tt}-\varphi_{,xx} =\sin\varphi,
    \end{equation}
    which got its name due to the similarity to the Klein-Gordon equation. It appears in in differential geometry and relativistic field theory. 
    
  \item The {\bf Kadomtsev--Petviashvili (KP) equation}
  \begin{equation}
   (u_{,t}+u_{,xxx}+6uu_{,x})_{,x}+3\sigma^2u_{,yy}=0.
  \end{equation}
   This is a generalisation of the KdV equation and describes water waves that depend on \emph{two} spacial coordinates $x$ and $y$.
   
  \item The {\bf nonlinear Schr\"odinger equation}
  \begin{equation}
   \ii\psi_{,t}+\frac{1}{2}\psi_{,xx}=\kappa|\psi|^2\psi,
  \end{equation}
  which has applications in quantum mechanics, optics and water waves.   
\end{itemize}

How is it possible that these equations can be studied in great detail, while general \emph{nonlinear} partial differential equations (PDEs) usually evade all attempts to construct (physically relevant) exact solutions? The reason is that those special equations have an underlying hidden structure: One can construct a corresponding  \emph{linear} matrix problem such that the nonlinear equation is equivalent to the integrability condition of the linear problem \cite{AKNS1974}. The linear problem has the form
\begin{equation}
  \Phi_{,x} = U\Phi,\quad
  \Phi_{,t} = V\Phi,
\end{equation}
 where the matrix function $\Phi=\Phi(x,t,\lambda)$ is the unknown, $U$ and $V$ are given matrix functions, and $\lambda$ is the so-called spectral parameter, an important additional degree of freedom. For example, if we choose
 \begin{equation}
  U=\begin{pmatrix}
      -\ii\lambda & u\\
      -1           & \ii\lambda
     \end{pmatrix},\quad
   V = \begin{pmatrix}
        -4\ii\lambda^3+2\ii u\lambda-p & 4u\lambda^2+2\ii p\lambda+s\\
        -4\lambda^2+2u & 4\ii\lambda^3-2\ii u\lambda+p
       \end{pmatrix},
 \end{equation}
then the functions $p(x,t)$ and $s(x,t)$ can be eliminated from the integrability condition $\Phi_{,xt}=\Phi_{,tx}$, which can also be expressed in terms of $U$ and $V$ as
$U_{,t}-V_{,x}+[U,V]=0$. The resulting condition is nothing but the KdV equation \eqref{eq:KdV} for the remaining function $u(x,t)$.

Once a linear problem is found for a nonlinear equation, various techniques from  soliton theory can be used to study the equation and construct exact solutions. These include B\"acklund transformations \cite{Baecklund1880, ZakharovShabat1974, ZakharovShabat1979} and Riemann--Hilbert problems \cite{Gardner1967}.

There is also an alternative approach to discuss soliton equations, which uses so-called Lax pairs \cite{Lax1968}. These are pairs of linear differential operators $P$ and $L$ satisfying Lax's equation
\begin{equation}
 \frac{\dd L}{\dd t}=[P,L]. 
\end{equation}
Here, we obtain the KdV equation if we choose $L=\partial_x^2+u$ and $P=-(4\partial_x^3+6u\partial_x+3 u_{,x})$.

In the following, we are particularly interested in a soliton equation that is relevant in the context of equilibrium configurations in general relativity. For that purpose, we  first summarise a few basics of relativity.

\subsection{General relativity and the black hole balance problem}

General relativity is Albert Einstein's geometric theory of gravitation, which was published in 1915 \cite{Einstein1915}. Space and time are unified into a 4-dimensional entity known as \emph{spacetime}, and the Newtonian gravitational force from classical mechanics is replaced by the concept of \emph{curvature}: Spacetimes are not flat, but curved. Consequently, particles travelling through a gravitational field are not moving on their particular trajectories because of a gravitational force, but simply due to the curvature of the surrounding spacetime.

Mathematically, a spacetime corresponds to a pair $(M,g)$, where $M$ is a Lorentzian manifold and $g$ a metric tensor, which determines distances, angles, curvature and particle paths. We can represent the metric by a symmetric $4\times 4$ matrix with components $g_{ij}$, $i,j=1,\dots,4$. Spatial length or time intervals between two events $x^i$ and $x^i+\dd x^i$ are then given in terms of the line element $\dd s^2=g_{ij}\,\dd x^{\,i}\dd x^{\,j}$, where we use Einstein's summation convention and sum over pairs of repeated indices.

The metric cannot be chosen arbitrarily. 
Instead, it is determined by the distribution of matter and energy in the spacetime and must satisfy \emph{Einstein's field equations}. In geometric units, where the speed of light and Newton's gravitational constant are equal to one, they read
\begin{equation}\label{eq:EinsteinEQ}
 G_{ij}=8\pi T_{ij}.
\end{equation}
The object on the left-hand side is the Einstein tensor, which is related to the curvature and can be obtained from $g_{ij}$ and its partial derivatives up to second order. The distribution of matter and any other fields is described by the energy-momentum tensor $T_{ij}$ on the right-hand side. 
Note that, throughout the article, we consider asymptotically flat spacetimes, which approach the flat Minkowski spacetime at infinity. As a consequence, a possible additional term $\Lambda g_{ij}$ in the Einstein equations is not present, corresponding to vanishing of the so-called cosmological constant $\Lambda$.

In general, \eqref{eq:EinsteinEQ} is a notoriously complicated system of ten coupled, nonlinear PDEs for the ten independent components of the metric $g_{ij}$. However, the equations become simpler when we consider spacetimes with symmetries, as we do in this paper, where we focus on \emph{axisymmetric and stationary} solutions. The resulting Einstein equations can then be reformulated in terms of a system of \emph{soliton} equations --- the Ernst equations. But before we look at these, we first explain the `black hole balance problem' in more detail.

In Newtonian physics, a configuration of several extended bodies separated by a plane clearly cannot be in equilibrium.\footnote{Note that the requirement of a separating plane is crucial here. Otherwise, one could have one large and strongly deformed body that almost encloses a smaller body, which can be arranged such that the configuration is in equilibrium.} 
Even if all bodies are initially at rest, they will accelerate towards the centre of gravity of the others and hence immediately start moving.
While this sounds trivial, it reveals an important property of Newtonian gravity: The gravitational interaction is never repulsive but always attractive. 

Could this change in the context of general relativity? If we consider \emph{static} configurations that are reflectionally symmetric with respect to a plane, then a result by Beig and Schoen \cite{BeigSchoen2009} shows that such static equilibrium configurations do not exist. However, what happens if we allow \emph{stationary} rather than static configurations, where axisymmetric bodies can \emph{rotate} about the symmetry axis? 
In the presence of several rotating objects, a new effect in general relativity is the spin-spin interaction, which is repulsive when two objects rotate in the same direction. 

In order to avoid complicated descriptions of matter distributions, it is natural to consider the probably simplest type of objects in general relativity: black holes (BHs). 
Although BHs are expected to contain singularities in their interior, they are very simple outside their event horizons (the BH surfaces), where we have nothing but vacuum in pure BH configurations.
This raises the question as to whether two or more aligned axisymmetric BHs can be in stationary equilibrium, which is the \emph{black hole balance problem}. 

Discussing this question is fundamentally important for understanding the nature of the gravitational interaction in general relativity. Moreover, it is crucial in the context of BH uniqueness: In our universe, we expect that any star that is heavy enough to collapse to a BH at the end of its life will eventually settle down to a Kerr BH --- a solution constructed in 1963 by Kerr \cite{Kerr1963}. 
Assuming that a solution to Einstein’s field equations contains a single BH (along with some additional mathematical conditions), it is known that the Kerr solution is the unique vacuum BH solution, see \cite{ChruscielCostaHeusler2012} and references therein.
However, as long as there is the possibility that a collapsing star could also form a stationary multi-BH solution, it cannot be absolutely guaranteed that the end product will always be a Kerr BH.

An interesting generalisation of this problem is the following. We could also allow the BHs to be \emph{charged}, in which case they would be surrounded by electrovacuum (i.e.\  vacuum plus electromagnetic fields). Due to the repulsion of like charges, such BHs may even have a better chance to form an equilibrium configuration.

At first glance, it might appear straightforward to construct an $n$-BH equilibrium configuration: We could simply choose BHs that are rotating fast enough and are sufficiently charged, so that the spin-spin and electromagnetic repulsions become strong enough to compensate the gravitational attraction. However, there is an upper limit for the charge and rotation rate beyond which BHs cannot exist. If we exceed this limit, then we obtain a solution that contains a singularity that is not surrounded by an event horizon. However, the presence of a horizon is the defining property for a BH. Objects without a horizon are called naked singularities, and they are considered unphysical and not expected to exist in our universe. Therefore, we need to restrict ourselves to \emph{subextremal} BHs with rotation rate and charge below the theoretical limit. 
We also exclude \emph{extremal} BHs, where the rates take on the largest possible values. While such BHs do have an event horizon, they are typically considered unphysical, as it is believed that they could only be approached but never exactly reached in nature, see, e.g.\ \cite{Thorne1974}.\footnote{If we allow extremal BHs, then an equilibrium configuration does exist and is given by the Majumdar--Papapetrou solution \cite{Majumdar1947, Papapetrou1947}. However, since this configuration requires extremal BHs, it is generally thought not to form in realistic astrophysical processes. Recent studies indicate, however, that in the presence of certain exotic types of matter, such as charged scalar fields or charged collisionless Vlasov matter, it may be possible for a subextremal black hole to evolve into an extremal one in finite time  \cite{KehleUnger2022, KehleUnger2024}. But so far there is no indication that the same would be possible in physically more realistic models.}

Finally, we reiterate that our considerations focus on asymptotically flat spacetimes with vanishing cosmological constant $\Lambda$. Otherwise, in the case $\Lambda>0$, static and stationary black hole binary configurations do interestingly exist, as the numerical constructions in \cite{DiasGibbonsSantosWay2023, DiasSantosWay2024} show. The black holes in those configurations are held apart by the cosmological expansion --- which, however, is not present in our setting with $\Lambda=0$.

In the following, we give the mathematical formulation of the BH balance problem and show how this naturally leads to a boundary value problem for a system of soliton equations. At present, this problem is unsolved in full generality, but we discuss known partial results.

\section{The Ernst equations and the linear problem}

We want to formulate a boundary value problem describing a stationary vacuum or electrovacuum spacetime with $n$ BHs. Some immediate simplifications are possible thanks to the following results. Firstly,  the event horizons of stationary BHs necessarily have spherical topology \cite{ChruscielWald1994}. (Note that this is a special property of our world with \emph{four} spacetime dimensions. If one formally considers BHs in higher dimensional manifolds, then they can have toroidal topology as well.) Secondly, our requirement of a stationary (time-independent) solution implies the existence of a second symmetry: The spacetime must be axisymmetric as well, i.e.\  independent of rotations about a symmetry axis, see \cite{Chrusciel1997} and references therein.

For an axisymmetric and stationary BH spacetime in electrovacuum, we can introduce cylindrical Weyl--Lewis--Papapetrou coordinates $\rho$, $\zeta$, $\varphi$, $t$ and write the line element as
\begin{equation}\label{eq:metric}
 \dd s^2 =f^{-1}[\ee^{2k}(\dd\rho^2+\dd\zeta^2)+\rho^2\dd\varphi^2]
            -f(\dd t+a\,\dd\varphi)^2
\end{equation}
in terms of three unknown functions $f=f(\rho,\zeta)$, $k=k(\rho,\zeta)$, $a=a(\rho,\zeta)$. 
These coordinates are adjusted to the symmetries of the solution, such that the metric functions do not depend on the symmetry variables $\varphi$ (axisymmetry) and $t$ (stationarity). See Sec.~13 in \cite{GriffithsPodolsky} and references therein for more details.

The Einstein--Maxwell equations for the (real) metric functions can be reformulated in form of the remarkably concise pair of \emph{Ernst equations} \cite{Ernst1968b}
\begin{equation}\label{eq:Ernst}
 f\Delta\E   = (\nabla\E+2\bar\Phi\nabla\Phi)\cdot\nabla\E,\quad
  f\Delta\Phi = (\nabla\E+2\bar\Phi\nabla\Phi)\cdot\nabla\Phi
\end{equation}
for the complex Ernst potentials $\E$ and $\Phi$, which encode the metric and electromagnetic degrees of freedom. Here, $\Delta$ and $\nabla$ are the Laplace and nabla operators in flat cylindrical coordinates, respectively, and $f$ is related to the Ernst potentials via $f=\Re(\E)+|\Phi|^2$.

The starting point for the derivation of the Ernst equations \eqref{eq:Ernst} is the observation that a part of the field equations for axisymmetric and stationary electrovacuum solutions can be solved by introducing suitable potential functions, which can be constructed from the components of the electromagnetic 4-potential and the metric functions $f$ and $a$. 
The real degrees of freedom are then combined into the two complex Ernst potentials $\E$ and $\Phi$. This allows us to express additional field equations, together with integrability conditions that need to be satisfied by the potential functions, in form of the Ernst equations. For more details on the Ernst equations for vacuum and electrovacuum solutions, including the explicit formulae for the Ernst potentials in terms of the metric and 4-potential, see Secs.~13.2 and 13.3 in \cite{GriffithsPodolsky} and references therein.

While \eqref{eq:Ernst} is a complicated system of coupled nonlinear PDEs, it belongs to the class of soliton equations and can be reformulated in terms of an associated linear problem \cite{Belinski1979, NeugebauerKramer1983, Meinel2012}. Indeed, \eqref{eq:Ernst} is equivalent to the integrability condition of the system
\begin{eqnarray}
 \Y_{,z} &= \left[
 \left(\begin{array}{ccc}
   B_1 & 0 & C_1\\ 0 & A_1 & 0\\ D_1 & 0 & 0
 \end{array}\right)
 +\lambda
 \left(\begin{array}{ccc}
  0 & B_1 & 0\\ A_1 & 0 & -C_1\\ 0 & D_1 & 0
 \end{array}\right)\right]\Y,\\
 \Y_{,\bar z} &= \left[
 \left(\begin{array}{ccc}
   B_2 & 0 & C_2\\ 0 & A_2 & 0\\ D_2 & 0 & 0
 \end{array}\right)
 +\displaystyle\frac{1}{\lambda}
 \left(\begin{array}{ccc}
  0 & B_2 & 0\\ A_2 & 0 & -C_2\\ 0 & D_2 & 0
 \end{array}\right)\right]\Y
\end{eqnarray}
for a $3\times3$ matrix function $\Y(z,\bar z;\lambda)$. Here, $z=\rho+\ii\zeta$ is a complex coordinate, and the matrix components are related to the Ernst potentials via
\begin{eqnarray}
\label{eq:LP1}
 &&A_1 = \bar B_2 = \frac{1}{2f}(\E_{,z}+2\bar\Phi\Phi_{,z}),\quad
 C_1 = f\bar D_2 = \Phi_{,z},\\
\label{eq:LP2}
 &&A_2 = \bar B_1 = \frac{1}{2f}(\E_{,\bar z}+2\bar\Phi\Phi_{,\bar z}),\quad
 C_2 = f\bar D_1 = \Phi_{,\bar z}.
\end{eqnarray}
Moreover, $\lambda$ is the square root function
\begin{equation}\label{eq:lambda}
 \lambda=\sqrt{\frac{K-\ii\bar z}{K+\ii z}}, 
\end{equation}
which depends on the coordinates and on a spectral parameter $K\in\CC$.

\begin{figure}\centering
 \includegraphics[width=7cm]{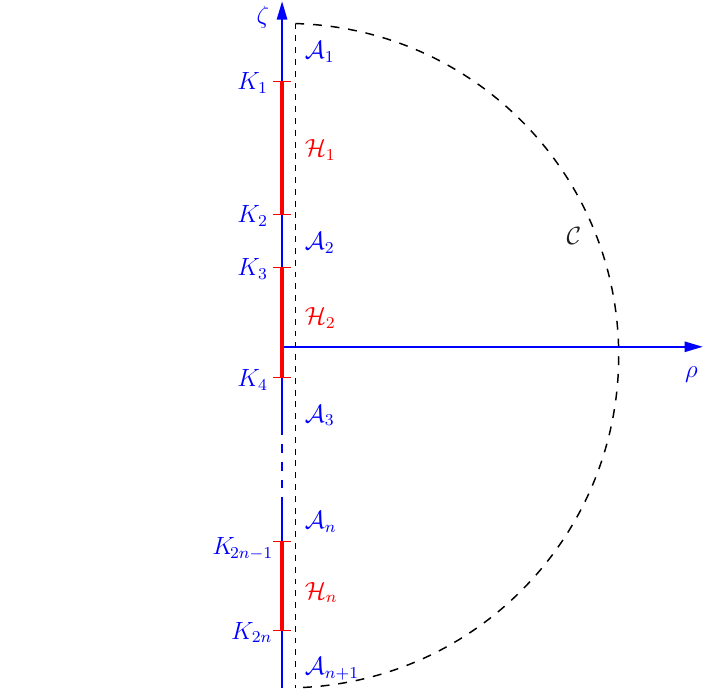}
 \caption{\label{fig:BVP}Illustration of the boundary value problem for an $n$-BH equilibrium configuration in Weyl--Lewis--Papapetrou coordinates.}
\end{figure}

A useful property of Weyl--Lewis--Papapetrou coordinates is that the BH horizons (which we know to have a spherical topology) degenerate into intervals on the $\zeta$-axis. Hence, this axis will contain $n$ intervals corresponding to the surfaces of the $n$ BHs in our (hypothetical) equilibrium configuration, while the remaining parts of the $\zeta$-axis represent the symmetry axis. We denote the horizons by $\mathcal H_1$, $\dots$, $\mathcal H_n$, the endpoints of the corresponding intervals (the north and south poles of the BHs) by $K_1$, $K_2$, $\dots$, $K_{2n-1}$, $K_{2n}$, and the parts of the symmetry axis by $\A_1$, $\dots$, $\A_{n+1}$, see Fig.~\ref{fig:BVP}. Note that our coordinates only cover the electrovacuum region \emph{outside} the BHs, i.e.\  we cannot (and do not need to!) access the interior of the BHs.

We now arrive at the following boundary value problem \cite{Hennig2020}: In the region $\rho\ge0$, $\zeta\in\R$, we want to solve the Ernst equations \eqref{eq:Ernst}, or, equivalently, the linear problem \eqref{eq:LP1}, \eqref{eq:LP2}.
At the boundaries, we impose the following conditions, which are consequences of the requirement of appropriate behaviour at horizons, symmetry axes and  infinity:
\begin{equation}\label{eq:boundary}
 \begin{aligned}
    &\mbox{horizons } \H_k:\quad     && a=-\frac{1}{\Omega_k},\\
    &\mbox{axes parts }\A_k:\quad && a=0,\\
    &\mbox{infinity }\C:          && \E\to 1,\ \Phi\to0,\\
    &\mbox{north/south poles $K_k$}:  && f=0.
   \end{aligned}
\end{equation}   
Here, $\Omega_k$ is the constant angular velocity of the $k$\textsuperscript{th} BH, and $\C$ is an infinitely large semicircle.   
      
In the next section, we will integrate the linear problem along the boundaries of the physical domain. This allows us to significantly restrict the mathematical form of the solutions describing BH equilibrium configurations, if any exist.

\section{The linear problem on the boundaries and calculation of the axis Ernst potentials}
\label{sec:LP}

In our discussion of the linear problem for the Ernst equations, it is important to observe that the solution $\Y$, as a function of the spectral parameter $K$ for fixed coordinates $z$, $\bar z$, is defined on a 2-sheeted Riemannian $K$-surface. This is because the function $\lambda$ [cf.~\eqref{eq:lambda}] in the linear problem, as a square root function, is naturally defined on 2-sheeted Riemannian surface. Only at the branch points $K_A:=\ii\bar z$ and $K_B:=-\ii z$, $\lambda$ is unique and takes on the values $0$ and $\infty$, respectively. Fortunately, it is not necessary to solve the linear problem on both sheets separately. Instead, once we know $\Y$ on a Riemannian sheet corresponding to some $\lambda$, we can easily construct the solution on the other sheet corresponding to $-\lambda$ by virtue of the transformation formula \cite{Hennig2019}
\begin{equation}\label{eq:sheets}
 \Y|_{-\lambda}=\J\Y|_{\lambda}\B(K),\quad
   \J=\mathrm{diag}(1,-1, 1),\quad
   \B=\B(K),\quad \B^2=\1.
\end{equation}
The $3\times3$ matrix function $\B$ could be completely fixed by imposing some gauge condition, but for our purposes it is sufficient to know that it satisfies $\B^2=\1$ (because, applying the transformation twice must give the identity transformation). 

Furthermore, since we consider BHs rotating with angular velocities $\Omega_k$, $k=1,\dots,n$, it is natural to also study the linear problem in a frame that corotates with one of the BHs. Coordinates adapted to the rotation of the $k$\textsuperscript{th} BH are easily obtained by changing the angle $\varphi$ to $\tilde\varphi=\varphi-\Omega_k t$.
The metric, the Ernst equation and the linear problem have the same mathematical form in the new coordinates, only the various functions change. Just like the transition from one Riemannian sheet to the other, the transformation to the new, rotating frame can be done with a useful formula. The solution $\tilde\Y$ in the rotating frame can be obtained from $\Y$ as follows \cite{AnsorgHennig2009, HennigAnsorg2009, Meinel2012},
\begin{equation}
 \tilde\Y=\left[
  \left(\begin{array}{ccc}
         c_- & 0  & 0 \\
         0   & c_+ & 0\\
         0   & 0   & 1
        \end{array}\right)
  +\ii(K+\ii z)\frac{\Omega_k}{f}
  \left(\begin{array}{ccc}
         -1      & -\lambda & 0\\
         \lambda & 1        & 0\\
         0       & 0        & 0
        \end{array}\right)\right]\Y,
\end{equation}
where $ c_{\pm}=1+\Omega_k\left(a\pm{\rho}/{f}\right)$. Since this transformation depends on the function $a$, it allows us to incorporate the correct boundary values \eqref{eq:boundary} at $\rho=0$.

We are now in a position to integrate the linear problem along the axes parts, the  horizons and the infinitely large semicircle $\C$, see Fig.~\ref{fig:BVP}.\footnote{Solving the linear problem for all values of $\rho$ and $\zeta$ appears unrealistic, since, despite being linear, the system \eqref{eq:LP1}, \eqref{eq:LP2} is still rather involved. However, the solution along the boundaries is completely sufficient and provides enough information for our purposes.}

Firstly, we consider the linear problem on the $\zeta$-axis ($\rho=0$). Here, we have $z=\ii\zeta=-\bar z$, and hence $\lambda$ simplifies to $\lambda=\pm1$ on the two Riemannian sheets.
The linear problem now reduces to a simple ODE, which we solve on the sheet with $\lambda=1$. [The solution on the other sheet quickly follows from \eqref{eq:sheets}.]

On the axis parts $\A_j$,  $j=1,\dots,n+1$, the solutions $\Y$ and $\tilde\Y$ in the nonrotating and corotating frames are
\begin{equation}\label{eq:axisY}
  \Y = \EE\Cmat_j,\quad
  \tilde\Y = \left[\EE+2\ii\Omega(K-\zeta)\left(\begin{array}{ccc}
                                                      -1 & 0 & 0\\
                                                      1 & 0 & 0\\
                                                      0 & 0 & 0
                                                     \end{array}\right)\right]\Cmat_j
\end{equation}
with `integration constants' $\Cmat_j$, which are $3\times 3$ matrices that are independent of the coordinates, but may depend on the spectral parameter $K$. Moreover, the matrix $\EE$ is given in terms of the axis values of the Ernst potentials,
\begin{equation}\label{eq:Ematrix}
 \EE:=\left(\begin{array}{ccc}
             \bar\E+2|\Phi|^2 & 1 & \Phi\\
             \E & -1 & -\Phi\\
             2\bar\Phi & 0 & 1
            \end{array}\right).
\end{equation}

Similarly, we obtain the solutions on the event horizons $\H_j$, $j=1,\dots,n$, in terms of additional integration constants $\Dmat_j$,
\begin{equation}
 \Y=\EE\Dmat_j,\quad
 \tilde\Y=\left[\left(\begin{array}{ccc}
                               0 & 0 & 0\\ 0 & 0 & 0\\ 0 & 0 & 1
                              \end{array}\right)
  \EE+2\ii\Omega(K-\zeta)\left(\begin{array}{ccc}
                                                      -1 & 0 & 0\\
                                                      1 & 0 & 0\\
                                                      0 & 0 & 0
                                                     \end{array}\right)\right]\Dmat_j.
\end{equation}

Secondly, we solve the linear problem at infinity, i.e.\  on the semicircle $\C$ in Fig.~\ref{fig:BVP}. Starting from a circle with finite radius $R$,  parametrised by 
$\rho=R\sin\alpha$, $\zeta=R\cos\alpha,\quad 0\le\alpha\le\pi$,
we find that in the limit $R\to\infty$, $\lambda$ simplifies to
$\lambda=\pm\ee^{\ii\alpha}$. Therefore, if we travel along $\C$, 
starting on the axis part $\A_1$ on the sheet with $\lambda=+1$, then the infinite semicircle leads us to the sheet with $\lambda=-1$ on $\A_{n+1}$, and vice versa. Consequently, we need to compare the $\lambda=1$ solution on $\A_1$ with the $\lambda=-1$ solution on $\A_{n+1}$. 
Moreover,  the linear problem simplifies to 
$\Y_{,z}=\0$, $\Y_{,\bar z}=\0$, i.e.\  $\Y$ is constant  on $\C$ (independent of $\rho$ and $\zeta$, but it could depend on $K$).

We have seen that the solution to the linear problem on the parts of the $\zeta$-axis as well as at infinity depends on various `integration constants' ($K$-dependent $3\times 3$ matrices). These, however, cannot be chosen independently. We have to impose the conditions of a continuous transition from one axis part to the next, i.e.\  continuity of both $\Y$ and $\tilde\Y$ at the north and south poles of the BHs. Moreover, we need to ensure the correct transition from $\A_1$ to $\A_{n+1}$ along $\C$, where, if we start on the sheet with $\lambda=1$, we will end up on the sheet with $\lambda=-1$. This gives the additional condition
\begin{equation}
 \lim_{\zeta\to\infty}\Y(0,\zeta;K)|_{\lambda=1}
    =\lim_{\zeta\to-\infty}\Y(0,\zeta;K)|_{\lambda=-1}.
\end{equation}
In this way, we obtain algebraic equations relating the matrices $\Cmat_j$ and $\Dmat_j$, which restrict the available degrees of freedom. The explicit continuity conditions can be found in \cite{Hennig2020}.

The information obtained about the solution $\Y$ in this way enables us to determine the Ernst potentials $\E$ and $\Phi$ on the $\zeta$-axis, where it is sufficient to consider $\A_1$, the uppermost part of the symmetry axis, see Fig.~\ref{fig:BVP}. In order to obtain the desired result, we come back to our previous observation that the solution $\Y$, for fixed coordinates and spectral parameter $K$, generally takes on different values on the `upper' and `lower' Riemannian sheet. Only at the two branch points $K_A=\ii\bar z$ and $K_B=-\ii z$, where $\lambda$ is unique, $\Y$ will take on the same value on both sheets, since they are connected there. If we approach the $\zeta$-axis, i.e.\  in the limit $\rho\to0$, both branch points come closer and closer. Finally, at $\rho=0$, we have confluent branch points at $K_A=K_B=\zeta$. 
The condition that $\Y$ for $\lambda=1$ and $\Y$ for $\lambda=-1$ on $\A_1$ coincide at $K=\zeta$ can be written as
\begin{equation}
 \EE\Cmat_1=\J\EE\Cmat_1\B  \quad\textrm{at}\quad K=\zeta.
\end{equation}

After eliminating $\B$ by virtue of the above continuity conditions, we obtain algebraic equations for the components of the matrix $\EE$, which are given in terms of the Ernst potentials, cf.~\eqref{eq:Ematrix}.
A careful analysis of these conditions reveals a special \emph{rational structure of the axis potentials}: The axis potentials for stationary $n$-black hole configurations, if any exist, necessarily have the form \cite{Hennig2020}
\begin{equation}\label{eq:genform}
 \A_1:\quad \E(0,\zeta)=\frac{\pi_n(\zeta)}{r_n(\zeta)},
      \quad \Phi(0,\zeta)=\frac{\pi_{n-1}(\zeta)}{r_n(\zeta)},
\end{equation}
where $\pi_n$ and $r_n$ are monic complex polynomials of degree $n$, and
$\pi_{n-1}$ is a complex polynomial of degree $n-1$. Therefore, the only remaining degrees of freedom are the corresponding polynomial coefficients.

Here, it is important to observe that the axis potentials on $\A_1$ already fix the solution uniquely for \emph{all} values of $\rho$ and $\zeta$ \cite{HauserErnst1981}. Consequently, once we have chosen the above polynomials, the solution is determined everywhere. In order to explicitly obtain the full exact solution, other methods from soliton theory can be employed, for example, `Sibgatullin's integral method' \cite{Sibgatullin1984, MankoSibgatullin1993}.

However, we stress that this result only provides a \emph{necessary} condition: If a regular equilibrium configuration with $n$ BHs exists, then the Ernst potentials are necessarily of the described form, i.e.\  they can be specified in terms of the polynomial coefficients that characterise the axis values on $\A_1$. 
On the other hand, if we arbitrarily choose some values for the polynomial coefficients, then we likely obtain only a formal, mathematical but \emph{unphysical} solution to Einstein's field equations, which needs to be dismissed. Physically acceptable solutions should satisfy all of the following additional conditions:
\begin{itemize}
 \item  The solutions have a vanishing NUT parameter, corresponding to
the appropriate behaviour at infinity in an asymptotically flat
spacetime. (The NUT parameter describes a geometric defect that can be interpreted as a type of gravitomagnetic monopole moment or a topological twist of spacetime. It was first introduced by Newman, Unti, and Tamburino \cite{NUT1963} for the Taub--NUT solution, which generalises the Schwarzschild black hole solution.)
\item There are no conical singularities on the symmetry axes, in
particular, between the BHs (which would correspond
to ‘struts’ that keep the BHs apart).
\item The norm of the axial Killing vector (a vector that mathematically describes the axisymmetry of the solution) vanishes on the symmetry
axis.
\item There is no global magnetic charge.
\item The solutions are free of singularities off the symmetry axis.
\end{itemize}

One can easily construct formal solutions to the BH balance problem that satisfy \emph{some} of these conditions. However, due to the complicated mathematical structure of those solutions, it is extremely hard to decide if there are any parameter choices for which \emph{all} above requirements are satisfied.

While the BH balance problem in its most general form (with an arbitrary number of BHs and including electromagnetic fields) is completely open, there are three interesting special cases, which we discuss in the next section.

\section{Known special cases}
\subsection{One black hole in vacuum}

We start with the simplest case $n=1$, where we only have a single BH. Although this does not actually describe a balance problem, it is still interesting to study the consequences of the above results in this situation. In other words, we want to obtain all vacuum spacetimes containing a single rotating BH. As mentioned in the introduction, it is known that the unique solution in this case must be the Kerr solution. However, the typical BH uniqueness proofs are quite abstract in nature. They show that, given the Kerr family of solutions, there cannot be solutions of any other form. This, however, requires that we already know the Kerr solution. 
From a physical perspective, it may be more satisfactory to formulate a boundary value problem describing the situation in question. By constructing all possible solutions, we can then observe that we obtain exactly the Kerr family of solutions.
This approach provides a \emph{constructive} uniqueness proof of the Kerr solution \cite{Neugebauer2000,NeugebauerMeinel2003}, which not only shows uniqueness, but we also find the explicit form of the solution.

In vacuum, we have $\Phi\equiv0$, and according to \eqref{eq:genform}, $\E$ has the form
\begin{equation}\label{eq:E1}
 \E(0,\zeta)=\frac{\zeta+c_1}{\zeta+c_2}
\end{equation}
on $\A_1$. The complex constants $c_1$, $c_2$ can be replaced in terms of physical parameters by comparing \eqref{eq:E1} with the general asymptotic behaviour of the Ernst potential $\E$ for $\rho=0$ and $\zeta\to\infty$, which is given by
\begin{equation}\label{eq:asympt}
 \Re(\E)=1-\frac{2M}{\zeta}+\mathcal O(\zeta^{-2}),\quad
   \Im(\E)=-\frac{2J}{\zeta^2}+\mathcal O(\zeta^{-3}).
\end{equation}
Here, $M$ is the total mass and $J$ the total angular momentum of the spacetime. 
Since our spacetime contains only a single BH, the mass and angular momentum are identical to those of the BH. Choosing the arbitrary $\zeta$-position of the BH such that the configuration is symmetric relatively to $\zeta=0$ (i.e.\  the north and south pole locations $K_1$ and $K_2$ satisfy $K_2=-K_1$) and using the continuity conditions, we can reformulate the axis Ernst potential \eqref{eq:E1} as
\begin{equation}
 \E(0,\zeta)=\frac{\zeta-M-\ii A}{\zeta+M-\ii A},
\end{equation}
where $A:=J/M$ is the rotation rate\footnote{The rotation rate of a Kerr BH is usually denoted by $a$, but since this symbol is already reserved for one of the metric functions in our line element \eqref{eq:metric}, we instead use $A$ here.}. 
This is indeed the known axis Ernst potential for the Kerr solution. Extending this solution to all space with other soliton methods, we eventually obtain the full Ernst potential of the Kerr solution,
\begin{equation}
 \E(\rho,\zeta)=\frac{R_+\ee^{-\ii\alpha}+R_-\ee^{\ii\alpha}-2M\cos\alpha}
         {R_+\ee^{-\ii\alpha}+R_-\ee^{\ii\alpha}+2M\cos\alpha}
\end{equation}
with
\begin{equation}
  R_\pm:=\sqrt{\rho^2+(\zeta\pm d)^2},\quad
  d=\sqrt{M^2-A^2},\quad
  A=M\sin\alpha.
\end{equation}
This completes the constructive uniqueness proof.

\subsection{One black hole in electrovacuum}

The Kerr BH has two parameters: the mass $M$ and the rotation rate $A$ (or, equivalently, the angular momentum $J$). If we allow the BH to have a charge $Q$ in addition, i.e.\  if we consider a rotating BH in \emph{electrovacuum}, then the Kerr solution generalises to the 3-parameter family of Kerr--Newman solutions \cite{Newman1965}. 
As with the Kerr solution in vacuum, the Kerr--Newman solution is unique among electrovacuum solutions containing a single BH. This can again be shown either with abstract methods \cite{Mazur1982, Bunting1983, Costa2010}, or one can give a constructive proof with the above-discussed soliton approach \cite{Meinel2012}.

Since the solution of the underlying boundary value problem with soliton methods is very similar to the discussion in the previous subsection, we do not go into any further details and just give the resulting Ernst potentials for the Kerr--Newman solution, 
\begin{equation}
 \E(r,\theta)    = 1-\frac{2M}{r-\ii A\cos\theta},\quad
 \Phi(r,\theta)  = \frac{Q}{r-\ii A\cos\theta}.
\end{equation}
We have conveniently expressed them in spherical Boyer--Lindquist coordinates $r$, $\theta$, which are related to our cylindrical coordinates $\rho$, $\zeta$ via
\begin{equation}
 \rho=\sqrt{r^2-2Mr+A^2+Q^2}\sin\theta,\quad
 \zeta=(r-M)\cos\theta.
\end{equation}

\subsection{Two black holes in vacuum}

Having established that the boundary value problem for a single BH ($n = 1$) can be completely solved, we now return to the BH balance problem, which requires at least two BHs. We next examine the case of two BHs ($n = 2$) in vacuum.
This is currently the only known situation where the BH balance problem has been resolved \cite{HennigNeugebauer2009, NeugebauerHennig2012, NeugebauerHennig2014, Chrusciel2011}. 

According to \eqref{eq:genform}, the corresponding axis potential $\E$ must be the quotient of two monic polynomials of second degree, while $\Phi$ vanishes in vacuum. This form of the axis potential uniquely characterises an already known exact solution: the \emph{double-Kerr--NUT solution} \cite{KramerNeugebauer1980}. Therefore, if any stationary binary BH configuration  existed, it would necessarily belong to this family of solutions. Originally, in \cite{KramerNeugebauer1980}, the Ernst potential for the double-Kerr--NUT solution was expressed as a quotient of two $5\times5$ determinants. Later, the following concise reformulation was discovered \cite{Yamazaki1983},
\begin{equation}
 \E(\rho,\zeta) = \frac{\left|\begin{array}{cc}
              R_{12}-1 & R_{14}-1\\
              R_{23}-1 & R_{34}-1\end{array}\right|}
              {\left|\begin{array}{cc}
              R_{12}+1 & R_{14}+1\\
              R_{23}+1 & R_{34}+1\end{array}\right|},
\end{equation}
where
\begin{equation}
 R_{ij}:=\frac{\alpha_ir_i-\alpha_jr_j}{K_i-K_j},\quad
 r_i:=\sqrt{\rho^2+(\zeta-K_i)^2},
\end{equation}
and the $\alpha_i$ are complex parameters with $|\alpha_i|=1$. The number of free parameters can be significantly reduced by imposing the first three regularity conditions from Sec.~\ref{sec:LP}. This leads to algebraic conditions that were explicitly solved in \cite{Manko2000, Manko2001}. 

The next difficult task is to ensure regularity off the symmetry axis (see the last condition in Sec.~\ref{sec:LP}). For any given choice of the remaining free parameters, we can plot the Ernst potential and observe that there are singularities, which suggests that there is no possible regular equilibrium configuration. However, the formula for the Ernst potential is  too complicated to rigorously prove this in full generality for any allowed parameter choice. 

Instead, we can use the following approach. In a stationary spacetime, the surface area $\mathcal A$ of the event horizon and the angular momentum $J$ of a subextremal BH cannot take on arbitrary values, but are restricted by the inequality \cite{HennigAnsorgCederbaum2008}
\begin{equation}\label{eq:inequality}
 8\pi|J|<\A.
\end{equation}
That is, for given `size' (area $\mathcal A$), there is an upper limit for the angular momentum.\footnote{A generalisation of this inequality can also be given for stationary \emph{charged} BHs  \cite{HennigCederbaumAnsorg2010}. Then the charge $Q$ enters the inequality, and we have $(8\pi J)^2+(4\pi Q^2)^2<\mathcal A^2$.} 

In order to decide if some of the remaining candidates for two-BH equilibrium configurations in the double-Kerr--NUT family of solutions are physically reasonable, we can study if there are choices of the parameters for which \emph{both} BHs satisfy the above inequality. 
Since the area $\mathcal A$ and the angular momentum $J$ can be read off from the solution at the event horizon, i.e.\  on the $\zeta$-axis, this analysis does not require to investigate the solution off the axis.

It turns out that, for any parameter choice, at least one of the two BHs violates \eqref{eq:inequality}. This shows that \emph{stationary equilibrium configurations with two aligned sub-extremal BHs in vacuum do not exist.}  For details of the proof, we refer to  \cite{HennigNeugebauer2009, NeugebauerHennig2012, NeugebauerHennig2014, Chrusciel2011}. 

\section{Discussion}

Soliton methods are extremely useful to study wave phenomena governed by integrable equations like the KdV equation, the KP equation, the Sine--Gordon equation or the nonlinear Schr\"odinger equation. 
However, soliton equations are also relevant in the context of stationary equilibrium configurations in general relativity. As an example, we have discussed the balance problem for axisymmetric rotating and possibly charged black holes. The question is whether the gravitational attraction can be balanced by the spin-spin and electromagnetic interactions. 

For two black holes in vacuum, a hypothetical equilibrium configuration needs to be in the class of double-Kerr--NUT solutions. However, all possible candidate solutions have been shown to be unphysical, proving that two black holes cannot be in stationary equilibrium. The problem remains open for two charged black holes in electrovacuum, as well as for three or more black holes in either vacuum or electrovacuum.
If such an equilibrium configuration existed, then it would necessarily fall into the class of solutions characterised by rational Ernst potentials on the symmetry axis.

We also considered the special case of just one black hole (which does not require any balancing). Using soliton methods, it is possible to obtain all corresponding single-black hole solutions, which provides constructive uniqueness proofs for the well-known  exact solutions for axisymmetric and stationary black holes, namely the Kerr solution in vacuum and the Kerr--Newman solution in electrovacuum.

Future research may uncover new $n$-black hole equilibrium configurations or show the non-existence thereof. This will further deepen our understanding of the intricate balance between gravitational, spin-spin, and electromagnetic interactions in these fascinating systems.


%


\end{document}